\documentclass[aps,showpacs,superscriptadress,preprint]{revtex4}
\usepackage{graphicx}

\begin{document}
\title{Consistent Thermodynamics for Quasiparticle Boson System with Zero Chemical Potential}

\author{Shaoyu Yin$^1$\footnote{051019008@fudan.edu.cn} and Ru-Keng
Su$^{1,2,3}$\footnote{rksu@fudan.ac.cn}} \affiliation{
\small 1. Department of Physics, Fudan University, Shanghai 200433, P. R. China\\
\small 2. CCAST(World Laboratory), P.O.Box 8730, Beijing 100080, P. R. China\\
\small 3. Center of Theoretical Nuclear Physics, National Laboratory
of Heavy Ion Collisions, Lanzhou 730000, P. R. China}

\begin{abstract}
The thermodynamic consistency of quasiparticle boson system with
effective mass $m^*$ and zero chemical potential is studied. We take
the quasiparticle gluon plasma model as a toy model. The failure of
previous treatments based on traditional partial derivative is
addressed. We show that a consistent thermodynamic treatment can be
applied to such boson system provided that a new degree of freedom
$m^*$ is introduced in the partial derivative calculation. A
pressure modification term different from the vacuum contribution is
derived based on the new independent variable $m^*$. A complete and
self-consistent thermodynamic treatment for quasiparticle system,
which can be widely applied to effective mass models, has been
constructed.
\end{abstract}


\maketitle

\section{Introduction}

System with complicated interaction is very common in real world and
rich in physical phenomenon but very hard to deal with. In practise,
physicists have introduced many theoretical tools, such as the
Hartree-Fock approximation, approximate secondary quantization,
summation of Feymann diagrams, \textit{etc.}, among which the
quasiparticle approximation is one of the simplest first order
approximation. Most of the quasiparticle approximations take the
system as consisting of quasiparticles without interaction, while
the effect of interaction is represented by the effective mass
$m^*$, usually being a function of temperature $T$ and particle
density $\rho$. It should be emphasized that the effective mass
quasiparticle model is only a rough approximation since many
detailed properties and higher order contributions are neglected.
However, to some degree, it is still an efficient tool to discuss
some macro properties of complex system because of its simplicity.
The application of quasiparticle model to the study of
thermodynamics can be found in many references, such as to the quark
system \cite{Fowler:1981,Chakrabarty:1989,Chakrabarty:dual,
Benrenuto:1995dual,Peng:1999,Wang:2000,Peng:2000,Zhang:2001,
Zhang:dual,Wen:2005} or quark-gluon plasma (QGP)
\cite{Goloviznin:1993,Peshier:1994,YangShinNan:1995,Peshier:1996,
Levai:1998,Schneider:2001,Biro:2003,Bannur:2007}.

However, it is well known that the effective mass quasiparticle
model suffers from the thermodynamic inconsistency. To illustrate
this inconsistency, for a quasiparticle boson system with chemical
potential $\mu=0$, the standard expressions of pressure,
$p=-\Omega/V=-\frac{gT}{2\pi^2}\int_0^\infty
k^2\ln[1-\exp(-\epsilon/T)]dk$, and internal energy density,
$\varepsilon=\frac{g}{2\pi^2}\int_0^\infty
k^2\frac{\epsilon}{\exp(\epsilon/T)-1}dk$, will not satisfy some
differential relations, such as $\varepsilon=T\frac{dp}{dT}-p$
\cite{YangShinNan:1995}. The problem comes from that the effective
mass, being a function of medium parameters, entangles the usual
thermodynamic relations via the relativistic dispersion relation
$\epsilon(k)=\sqrt{k^2+m^{*2}(T,\rho)}$. As was shown in
Ref.\cite{Yin:2008}, the usual thermodynamic treatment with
traditional partial derivative to obtain the thermodynamic
quantities is problematic. The reason is that the thermodynamic
potential $\Omega$ of quasiparticle model is not only a function of
characteristic variables $T$, $V$ and $\mu$, but also depends
explicitly on the effective mass, $\Omega=\Omega(T,V,\mu,m^*)$. To
overcome this difficulty, there are many different treatments in the
market. For example: (I). Calculate the derivatives of $m^*$ with
respect to $T$ and $\rho$ in the differential relations for a
reversible process,
\begin{equation}
d\Omega=-SdT-pdV-\overline{N}d\mu,
\end{equation}
\begin{equation}
S=-\left(\frac{\partial\Omega}{\partial T}\right)_{V,\mu},\qquad
p=-\left(\frac{\partial\Omega}{\partial V}\right)_{T,\mu},\qquad
\overline{N}=-\left(\frac{\partial\Omega}{\partial\mu}\right)_{T,V},
\end{equation}
where $S$, $p$ and $\overline{N}$ are entropy, pressure and average
particle number, respectively. Many extra terms involving
$\frac{\partial\Omega}{\partial m^*}\frac{\partial
m^*}{\partial\rho}$ and $\frac{\partial\Omega}{\partial
m^*}\frac{\partial m^*}{\partial T}$ will emerge in the expressions
of $p$, $\varepsilon$ and entropy density $s$
\cite{Benrenuto:1995dual,Peng:2000,Wen:2005}. But as was summarized
in Ref.\cite{Yin:2008}, though many authors follow this direction,
the extra terms in different references contradict each other. (II).
Introduce a temperature- and/or density-dependent vacuum energy
\cite{YangShinNan:1995,Biro:2003,Wang:2000} and force the vacuum to
cancel the thermodynamic inconsistency. However, the physics of the
vacuum is still unclear. (III). A new treatment had been suggested
in our previous paper \cite{Yin:2008}. We introduce a new degree of
freedom $m^*$ for the quasiparticle model effectively in the
thermodynamic derivative relation and rewrite Eq.(1) as
\begin{equation}
d\Omega=-SdT-pdV-\overline{N}d\mu+Xdm^*,
\end{equation}
then Eq.(2) becomes
\begin{equation}
S=-\left(\frac{\partial\Omega}{\partial T}\right)_{V,\mu,m^*},\quad
p=-\left(\frac{\partial\Omega}{\partial V}\right)_{T,\mu,m^*},\quad
\overline{N}=-\left(\frac{\partial\Omega}{\partial\mu}\right)_{T,V,m^*},
\quad X=\left(\frac{\partial\Omega}{\partial m^*}\right)_{T,V,\mu},
\end{equation}
where $X$ is an extensive quantity corresponding to the intensive
variable $m^*$. We emphasize that after introducing the term $Xdm^*$
in Eq.(3), we have Eq.(4) instead of Eq.(2). Since $m^*$ is an
invariant in the partial derivatives for $S$, $p$ and
$\overline{N}$, all the extra terms involving
$\frac{\partial\Omega}{\partial m^*}\frac{\partial
m^*}{\partial\rho}$ and $\frac{\partial\Omega}{\partial
m^*}\frac{\partial m^*}{\partial T}$ are forbidden there. We have
proven in Ref.\cite{Yin:2008}, by keeping $m^*$ invariant during the
derivatives, the results remain self-consistent and in accord with
the equilibrium statistics.

Noticing that at a fixed instant in the reversible process, the
system is in an equilibrium state with temperature and density
denoted as $T_0$ and $\rho_0$, respectively, then the effective mass
of the quasiparticle becomes constant $m^*(T_0,\rho_0)\equiv m_0$.
In this equilibrium state, the system reduces to a usual ideal gas
system with constant particle mass $m_0$, whose thermodynamic
quantities can be directly obtained statistically. In
Ref.\cite{Yin:2008} we have proven that the thermodynamic quantities
calculated in equilibrium state are just the corresponding
quantities given by Eqs.(3) and (4). The introduction of the new
variable $m^*$ can be understood in the following aspect. In the
quasiparticle approximation, the effective mass summarizes the
interaction, the confinement mechanism, \textit{etc.}, which must be
reflected in a consistent thermodynamic treatment. It forces us to
introduce $m^*$ as a new independent variable in the partial
derivative to get a suitable treatment. Employing the quark mass
density-dependent model, we had shown this treatment is
thermodynamically self-consistent \cite{Yin:2008}. The ambiguities
and inconsistencies in previous thermodynamic treatments are
overcome.

The motivation of this paper is to extend our study from the fermion
system to the boson system with non-conserved particles. The
chemical potential $\mu$ equals zero for such system. We here take a
quasiparticle gluon plasma (qGP) model, which was first introduced
in Refs.\cite{Goloviznin:1993,Peshier:1994}, as a toy model to
illustrate our treatment. This model is quite rough: On the one
hand, it can not describe the properties of the real QGP exhibited
by recent RHIC experiments \cite{Adams:2005,Adcox:2005}. One the
other hand, one can calculate the hard thermal or dense loops
\cite{Braaten:1990,Frenkel:1990,Blaizot:dual} by temperature quantum
theory and get higher order contributions beyond mean field and
quasiparticle picture. However, we will still employ this model. The
reasons are threefold: (1). The model is very simple and we can use
this model to illustrate our treatment transparently. (2). This
model has been employed by many authors to discuss the thermodynamic
inconsistency in the physics market \cite{YangShinNan:1995,
Biro:2003,Bannur:2007}. We use this model in favor of comparing our
treatment with others. (3). This model has the lattice simulation
results \cite{Engels:1989,Boyd:1995dual}, which can be used as
standards in comparing different treatments to expose their
advantages and shortcomings. We will focus our attention to
treatment (I) and treatment (III) since the study of qGP model by
treatment (II) was shown in Ref.\cite{YangShinNan:1995} in detail.
We will prove that the results given by traditional partial
derivative treatment (I) are completely different from those of the
lattice simulation. This treatment is failure for the qGP system as
for the quark system \cite{Yin:2008}.

Another important problem involves the negative pressure. Of course,
a real gluon plasma has no negative pressure. But as is predicted by
lattice simulation, the QGP system experiences a phase transition at
a critical temperature $T_c$ below which the system becomes
comfined. One need to introduce the negative pressure at $T<T_c$
indicating the instability in the quasiparticle model to describe
the phase transition at $T_c$. In treatment (I)
\cite{Benrenuto:1995dual, Peng:2000,Wen:2005}, the authors claimed
that the term $\frac{\partial\Omega}{\partial m^*}\frac{\partial
m^*}{\partial\rho}$ contributes a negative pressure which can cause
the instability at low $T$. But unluckily, in Ref.\cite{Yin:2008},
we proved this term can not exist in a consistent treatment. In
treatment (II), the authors obtained the negative pressure from the
vacuum \cite{YangShinNan:1995,Biro:2003,Wang:2000}. In this paper,
we will show that the negative pressure can exist naturally in
treatment (III). It is not necessary to involve the vacuum. The
newly independent variable $m^*$ permits a freedom for an arbitrary
function $f(m^*)$ in the formula of pressure, whose value can be
determined by the comparison with experiment results or lattice
data.

The paper is organized as follows: In the next section, using the
lattice data as the criterion for comparison, we will examine the
qGP model by treatment (I). In Sec. III, similar comparison will be
made between the lattice data and our treatment (III). Then in Sec.
IV, a modification term to pressure and entropy is derived within
our self-consistent treatment (III). We will show this modification
term is different from the vacuum contribution but a natural result
of our newly introduced variable $m^*$. Sec. V is a brief summary.

\section{The results of treatment (I)}

For the thermodynamic system with effective mass, after calculating
the derivatives of $m^*$ to $T$ and $\rho$ in Eq.(2), one obtained
\cite{Wen:2005}
\begin{eqnarray}
p&=&-\widetilde{\Omega}-V\frac{\partial\widetilde{\Omega}}{\partial
V}+\rho\frac{\partial\widetilde{\Omega}}{\partial
m^*}\frac{\partial m^*}{\partial\rho},\\
\varepsilon&=&\widetilde{\Omega}-\mu
\frac{\partial\widetilde{\Omega}}{\partial\mu}-
T\frac{\partial\widetilde{\Omega}}{\partial
T}-T\frac{\partial\widetilde{\Omega}} {\partial
m^*}\frac{\partial m^*}{\partial T},\\
s&=&-\frac{\partial\widetilde{\Omega}}{\partial T}-
\frac{\partial\widetilde{\Omega}}{\partial m^*}\frac{\partial
m^*}{\partial T},
\end{eqnarray}
where $\widetilde{\Omega}\equiv\Omega/V$ denotes the thermodynamic
potential density. The qGP model is a perfect boson system with
vanishing chemical potential $\mu$ and temperature- and
density-dependent particle mass $m^*$, so the thermodynamic
potential density
\begin{equation}
\widetilde{\Omega}=\frac{gT}{(2\pi)^3}\int\ln\left(1-
e^{-\beta\sqrt{k^2+m^{*2}}}\right)dk^3,
\end{equation}
which is explicitly independent of $V$ for infinite qGP system.
Eqs.(5) and (6) reduce to
\begin{eqnarray}
p&=&-\widetilde{\Omega}+\rho
\frac{\partial\widetilde{\Omega}}{\partial
m^*}\frac{\partial m^*}{\partial\rho},\\
\varepsilon&=&\widetilde{\Omega}-
T\frac{\partial\widetilde{\Omega}}{\partial
T}-T\frac{\partial\widetilde{\Omega}}{\partial m^*}\frac{\partial
m^*}{\partial T}.
\end{eqnarray}
Now we employ Eqs.(7-10) and three different models to study the
properties of qGP model. The results are shown in Figs.1-3.

The first model is an SU(2) qGP model. The detail of this model has
been addressed in Ref.\cite{YangShinNan:1995} and all the parameters
including the fitting of $m^*$ which we used for calculation are
from there. The results are shown in Fig.1, where the curves of
energy density, pressure and entropy density are normalized to their
Steven-Boltzmann limits, respectively. The solid curves adopted from
Ref.\cite{Engels:1989} are given by the lattice QCD calculation
which we use as the standard to compare different treatments. The
dotted curves are the results of Eqs.(7-10). We see from Fig.1 that
the deviation between solid curves and the corresponding dotted
curves are remarkable, especially in low temperature regions.

The second and the third models are SU(3) qGP models. The details of
these models have been addressed in Ref.\cite{Bannur:2007}, where
the effective mass of SU(3) gluon is given by
\begin{equation}
m^{*2}=\frac{a_0\alpha_s(T)\rho}{T},\quad\textrm{model I},
\end{equation}
and
\begin{equation}
m^{*2}=\frac{3a_0\alpha_s(T)\rho}{2T},\quad\textrm{model II},
\end{equation}
where the constant $a_0=2.15$ and the two-loop order running
coupling constant
\begin{equation}
\alpha_s(T)=\frac{2\pi}{11\ln(T/\Lambda_T)}
\left(1-\frac{51\ln(2\ln(T/\Lambda_T))}{121\ln(T/\Lambda_T)}\right).
\end{equation}
The only parameter $\lambda=\Lambda_T/T_c$ is fitted according to
the lattice data as $\lambda=0.83$ for model I and $\lambda=0.7$ for
model II. The results of treatment (I) are shown in Figs.2 and 3 for
model I and model II, respectively. The inconsistency is
transparently seen again. One can prove the deviation can not be
conciliated by different choices of $\lambda$. From Figs.1-3, the
failure of treatment (I) is obvious. The shapes of the curves are
completely different.

\section{The results of treatment (III)}

As demonstrated in our previous work \cite{Yin:2008}, the exact
differential relations of thermodynamic functions in quasiparticle
model should take the effective mass as a new independent degree of
freedom. In grand canonical ensemble, the exact differential
relation of $\Omega$ should be rewritten as Eqs.(3) and (4). We have
proven that these formulae are self-consistent and agree with the
statistical definitions and relations of thermodynamic quantities
\cite{Yin:2008}. Employing Eq.(4), Eq.(8) and the energy definition
\begin{equation}
U=\sum_ig_in_i\epsilon_i=\sum_i\frac{g_i\sqrt{m^{*2}+k_i^2}}{e^{\beta\sqrt{m^{*2}+k_i^2}}-1},\\
\end{equation}
we calculate the corresponding thermodynamic quantities by treatment
(III) for the SU(2) and SU(3) models in Sec. II. Our results are
shown in Figs.4-6. In Fig.4, the solid curves and the dashed curves
refer to the lattice data and the results of treatment (III) for the
SU(2) model, respectively. The solid curve and dashed curve for the
entropy density coincide since the function $m^*(T)$ are determined
from the lattice data of entropy \cite{YangShinNan:1995}. Figs.5 and
6 are the same as those of Figs.2 and 3, except that the dashed
curves are the results of our treatment (III). Comparing Figs.1-3
from treatment (I) with the corresponding Figs.4-6 from treatment
(III), we see that the curves of our treatment (III) lies much
closer to the lattice data, and the catastrophic deviation in
Figs.1-3 disappears.

We are convinced by these results as well as Ref.\cite{Yin:2008}
that treatment (I) with traditional partial derivative fails in the
quasiparticle model with effective mass, and the thermodynamically
consistent treatment (III), by introducing the new independent
variable $m^*$ in the partial derivative calculation, produces much
better results. However, treatment (III) in its present form can not
give some expected properties. We see from Figs.5 and 6, the
vanishing pressure at $T=T_c$ is exhibited by the lattice data,
since below $T_c$ the system becomes confined and no free particle
exists any more. But the present result of pressure of treatment
(III) for quasiparticle model is always finitely positive at $T>0$.
It is challenging to find a reasonable way to obtain a negative
pressure modification to realize the phase transition at $T_c$
within our consistent treatment, this will be the topic in next
section.

\section{Modification to pressure and entropy without vacuum}

Starting from the thermodynamic formula
\begin{equation}
\varepsilon+p-Ts=\frac{G}{V}=n\mu=0,
\end{equation}
for qGP model and Eq.(4), we get
\begin{equation}
\varepsilon=T\left(\frac{\partial p}{\partial
T}\right)_{m^*}-p=T^2\left(\frac{\partial(p/T)}{\partial
T}\right)_{m^*},
\end{equation}
The difference between Eq.(16) and its usual form,
$\varepsilon=T\frac{dp}{dT}-p$, is significant. Instead of the total
derivative $\frac{dp}{dT}$, we have the partial derivative with
$m^*$ fixed, because $m^*$ is an independent variable in Eq.(3). We
can integrate Eq.(16) and get
\begin{equation}
p=T\left[\int^T\frac{\varepsilon}{t^2}dt+f(m^*)\right],
\end{equation}
where $f(m^*)$ is an arbitrary integral function of $m^*$. Eq.(17)
shows the significant difference from the usual relation without
taking $m^*$ as a free parameter. Obviously, if we choose
$f(m^*)=0$, Eq.(17) reduces to the normal ideal gas case.

According to Eq.(17), the thermodynamic potential $\Omega$, entropy
$S$ and $X$ become
\begin{eqnarray}
\Omega&=&-pV=-TV\int^T\frac{\varepsilon}{t^2}dt-TVf(m^*),\\
S&=&-\left(\frac{\partial\Omega}{\partial T}\right)_{V,m^*}
=V\left[\frac{\varepsilon}{T}+\int^T\frac{\varepsilon}{t^2}dt
+f(m^*)\right],\\
X&=&\left(\frac{\partial\Omega}{\partial m^*}\right)_{T,V}
=-TV\int^T\frac{1}{t^2}\left(\frac{\partial\varepsilon}{\partial
m^*}\right)_{T,V}dt-TVf'(m^*),
\end{eqnarray}
respectively. Then, noticing that $\mu=0$, the differential relation
of internal energy $U$ reads
\begin{eqnarray}
dU&=&d(\Omega+TS)\nonumber\\
&=&TdS-pdV+Xdm^*\nonumber\\
&=&Td\left\{V\left[\frac{\varepsilon}{T}
+\int^T\frac{\varepsilon}{t^2}dt+f(m^*)\right]\right\}
-\left[T\int^T\frac{\varepsilon}{t^2}dt+Tf(m^*)\right]dV\nonumber\\
&&-\left[TV\int^T\frac{1}{t^2}\left(\frac{\partial\varepsilon}
{\partial m^*}\right)_{T,V}dt+TVf'(m^*)\right]dm^*\nonumber\\
&=&\left[\varepsilon+T\int^T\frac{\varepsilon}{t^2}dt+Tf(m^*)\right]dV\nonumber\\
&&+TV\left[\frac{d\varepsilon}{T}-\frac{\varepsilon}{T^2}dT
+\frac{\varepsilon}{T^2}dT+\int^T\frac{1}{t^2}
\left(\frac{\partial\varepsilon}{\partial
m^*}\right)_{T,V}dm^*dt+f'(m^*)dm^*\right]\nonumber\\
&&-\left[T\int^T\frac{\varepsilon}{t^2}dt+Tf(m^*)\right]dV
-TV\left[\int^T\frac{1}{t^2}\left(\frac{\partial\varepsilon}
{\partial m^*}\right)_{T,V}dtdm^*+f'(m^*)dm^*\right]\nonumber\\
&=&\varepsilon dV+Vd\varepsilon=d(V\varepsilon),
\end{eqnarray}
which means that $U$ is not affected by the arbitrary function
$f(m^*)$, and it still consists with the definition of internal
energy in Eq.(14). This result is quite reasonable for the physical
picture of the quasiparticle model. From Eqs. (17), (19) and (21),
the relation
\begin{equation}
S=\frac{U+pV}{T}
\end{equation}
remains correct.

Now we are in a position to determine $f(m^*)$ from the lattice
simulation data. Hereafter we choose the SU(3) model I and model II
as our working examples, taking advantage of the clearly expressed
$m^*$ in Eqs.(11) and (12).

With the effective mass in hand, the internal energy can be directly
calculated from Eq.(14). For the pressure and entropy, one first
calculate in equilibrium state as usual, the difference between the
results and the lattice data gives the unknown function $f(m^*)$, as
is shown in Eqs.(18) and (19). But since the internal energy
obtained by the given $m^*$ may generally different from the lattice
data, remembering $U=TS-pV$ is satisfied by both the theoretical
result and the data, the $f(m^*)$ fitted from the pressure and the
one from the entropy will be different, as was shown in Figs.7 and
8. Fig.7 refers to the SU(3) model I, where the dashed curve is for
the $f(m^*)/T^3$ fitted from the entropy data and the dotted curve
for the one fitted from the pressure data, respectively. Fig.8 is
the same as Fig.7 but for the SU(3) model II. We see from these two
figures that the functions $f(m^*)$ obtained from the lattice
pressure data and the entropy data lie quite close to each other.
With these results, the modified pressure and entropy density for
these two models are shown in Figs.9 and 10, with the dotted line
representing the entropy density modified by $f(m^*)$ fitted from
the pressure; and the dashed line representing the pressure modified
by $f(m^*)$ fitted from the entropy, respectively. We see from these
two figures that the curves of modified pressure and entropy density
agree with the lattice data quite well. The internal energy density
is not changed due to Eq.(21), so it remains the same as in Figs.5
and 6 and is not plotted again in Figs.9 and 10. It is worth
pointing out that if $m^*$ is not given previously but fitted by the
internal energy from Eq.(14), the $f(m^*)$ fitted from the pressure
and the one from the entropy will be the same, and all the three
curves will coincide exactly with the lattice data. In particular,
the segments of negative pressure are plotted explicitly below $T_c$
in Figs.9 and 10. The function $f(m^*)$ contributes the negative
pressure in a natural fashion. The appearance of the negative
pressure, unlike treatment (II) \cite{YangShinNan:1995,
Biro:2003,Wang:2000}, does not relies on the vacuum. This is an
important character of our treatment (III). At last, we would like
to emphasize again that the negative pressure shown in Figs.9 and 10
at $T<T_c$ does not means such states really exist. In fact the
quasiparticle model is only applicable for $T>T_c$. We keep the
segments of the curves below $T_c$ just to demonstrate explicitly
how the mechanism of negative pressure instability acts in the
quasiparticle model within our treatment.

\section{Summary}

With the qGP model as a toy model, we demonstrate explicitly again
the difficulty of treatment (I) with traditional partial derivative
and prove that our treatment (III) is applicable, by the criterion
of the lattice data. We also demonstrate the success of the newly
introduced function $f(m^*)$ which can fit the lattice data
self-consistently. It's worth emphasizing that the physical meaning
of the additional function is completely different from the vacuum
energy. This function does not appear in the expression of
$\varepsilon$, but change the values of $p$ and $s$. The basic
differences between our treatment (III) and other treatments are as
follows: (1). As was proven in Ref.\cite{Yin:2008}, treatment (III)
is thermodynamically self-consistent and coincide with the
statistics in equilibrium state. We do not need to introduce a
vacuum and force it to cancel the thermodynamically inconsistent
terms as in treatment (II). (2). We have an vacuum-unrelated
integral function $f(m^*)$ in the consistent treatment, which can be
fixed by fitting to experiment data or standard theoretical results.

Together with our previous paper \cite{Yin:2008}, we have
established a self-consistent and complete thermodynamics for the
effective mass quasiparticle model. Our treatment (III) is not only
applicable to fermion system with nonzero chemical potential, but
also to boson system with zero chemical potential. It can be wildly
applied in all quasiparticle system.

\section*{Acknowledgements}

This work is supported in part by NNSF of China. S. Yin is partially
supported by the graduate renovation foundation of Fudan university.

\begin{figure}[tbp]
\includegraphics[totalheight=16cm, width=16cm]{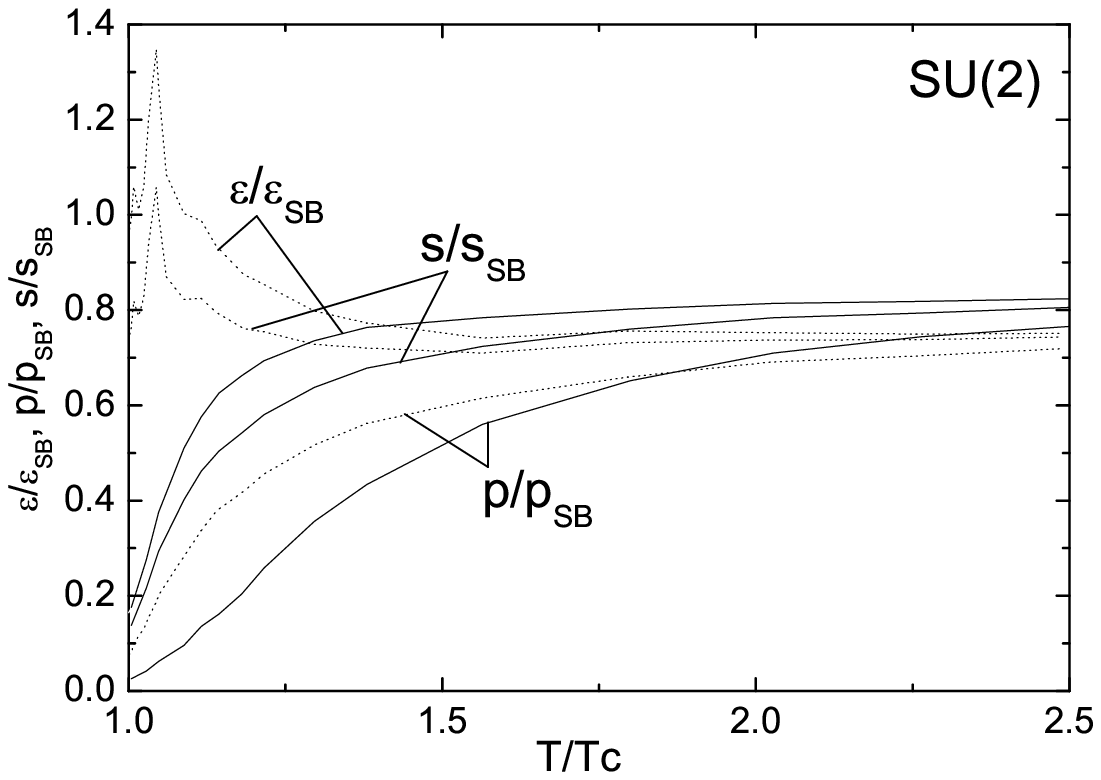}
\caption{Internal energy density, pressure and entropy density as
functions of temperature in SU(2) GP system. The solid curves are
the lattice data from Ref.\cite{Engels:1989}; while the dotted
curves are the results of treatment (I).} \label{fig1}
\end{figure}

\begin{figure}[tbp]
\includegraphics[totalheight=16cm, width=16cm]{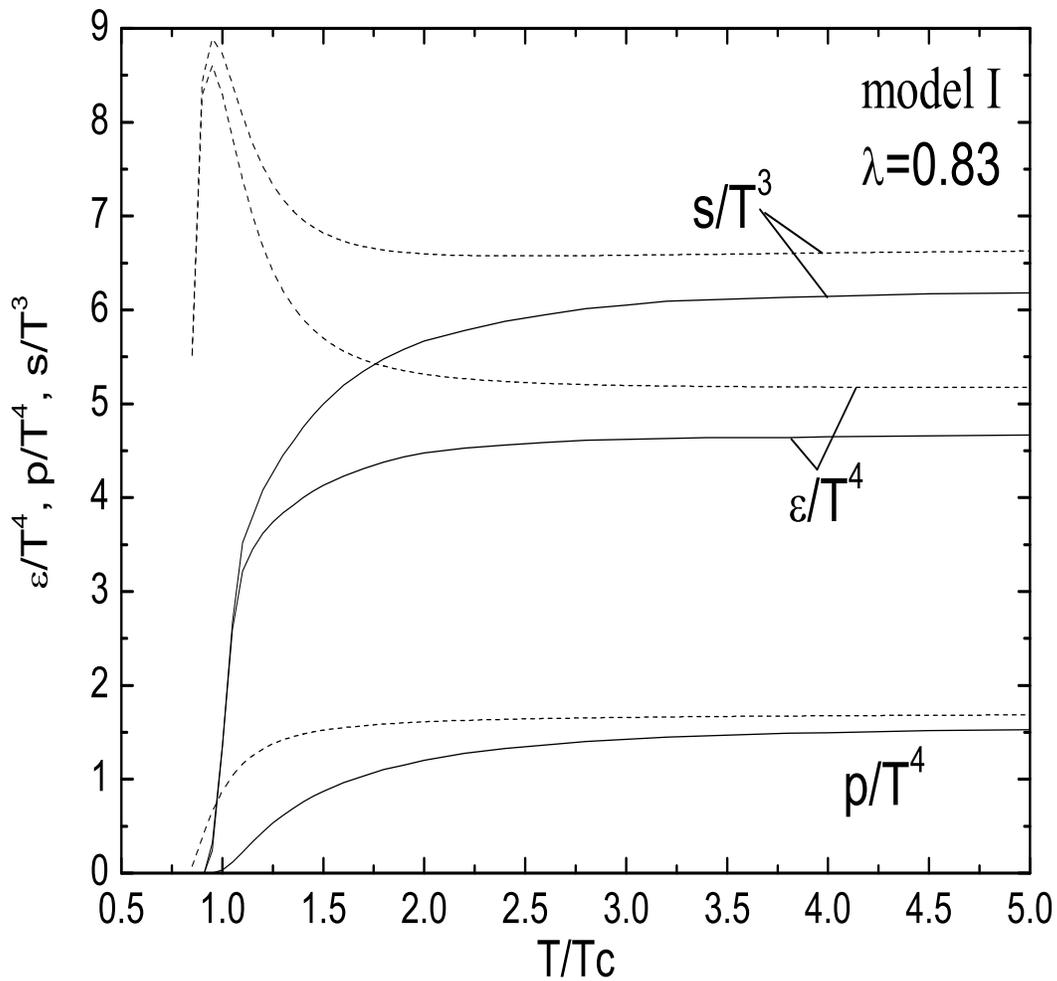}
\caption{Internal energy density, pressure and entropy density as
functions of temperature in SU(3) GP system with effective mass of
model I and $\lambda=0.83$. The solid curves are the lattice
simulation data from Ref.\cite{Boyd:1995dual}, and dashed curves are
the results of treatment (I).} \label{fig2}
\end{figure}

\begin{figure}[tbp]
\includegraphics[totalheight=16cm, width=16cm]{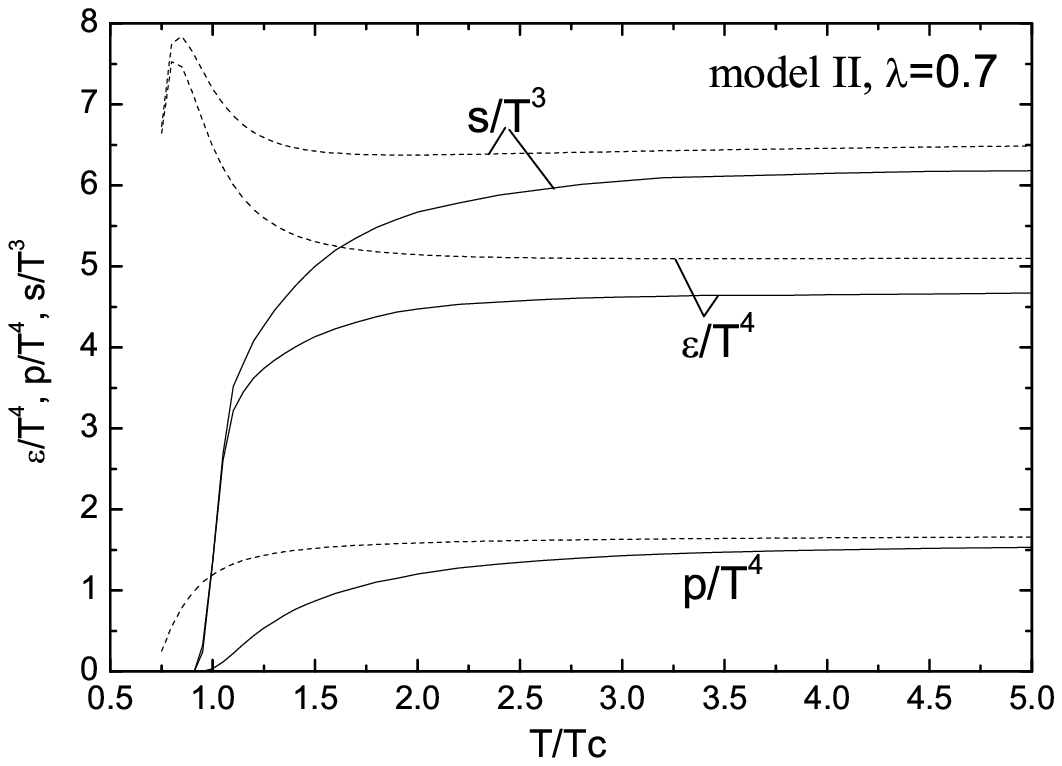}
\caption{The same as Fig.2, except that the effective mass is of
model II, and $\lambda=0.7$.} \label{fig3}
\end{figure}

\begin{figure}[tbp]
\includegraphics[totalheight=16cm, width=16cm]{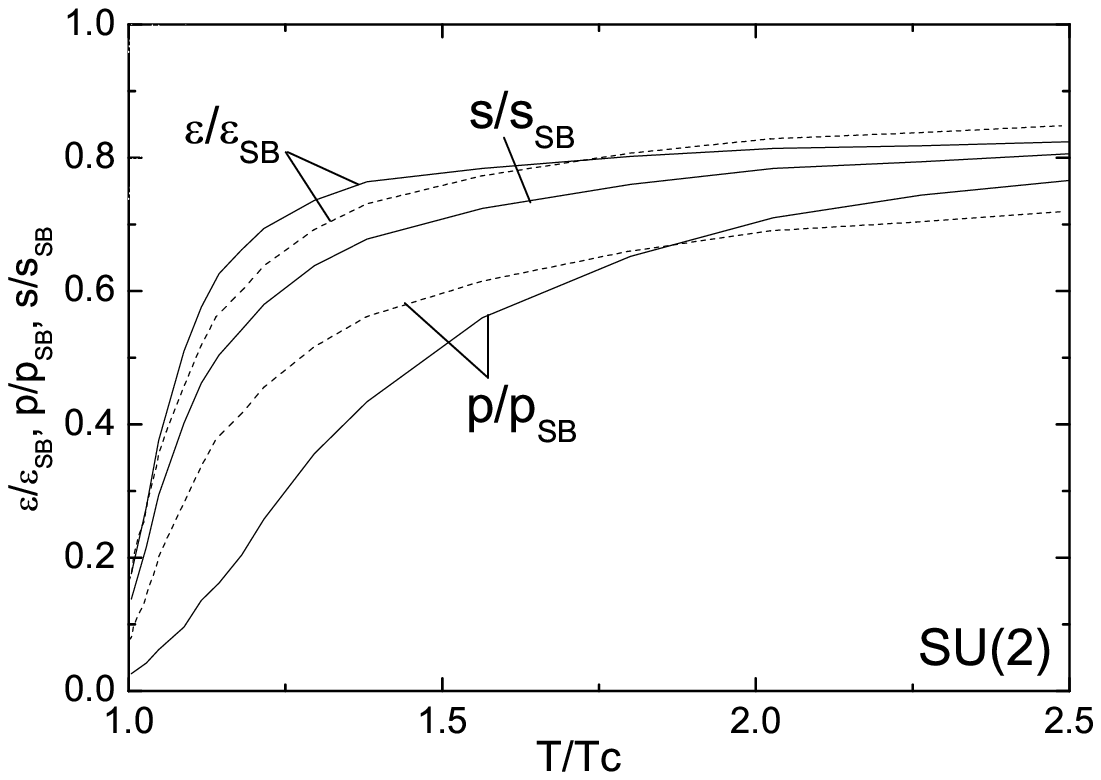}
\caption{The same as Fig.1, except that the dashed curves are the
results of treatment (III).} \label{fig4}
\end{figure}

\begin{figure}[tbp]
\includegraphics[totalheight=16cm, width=16cm]{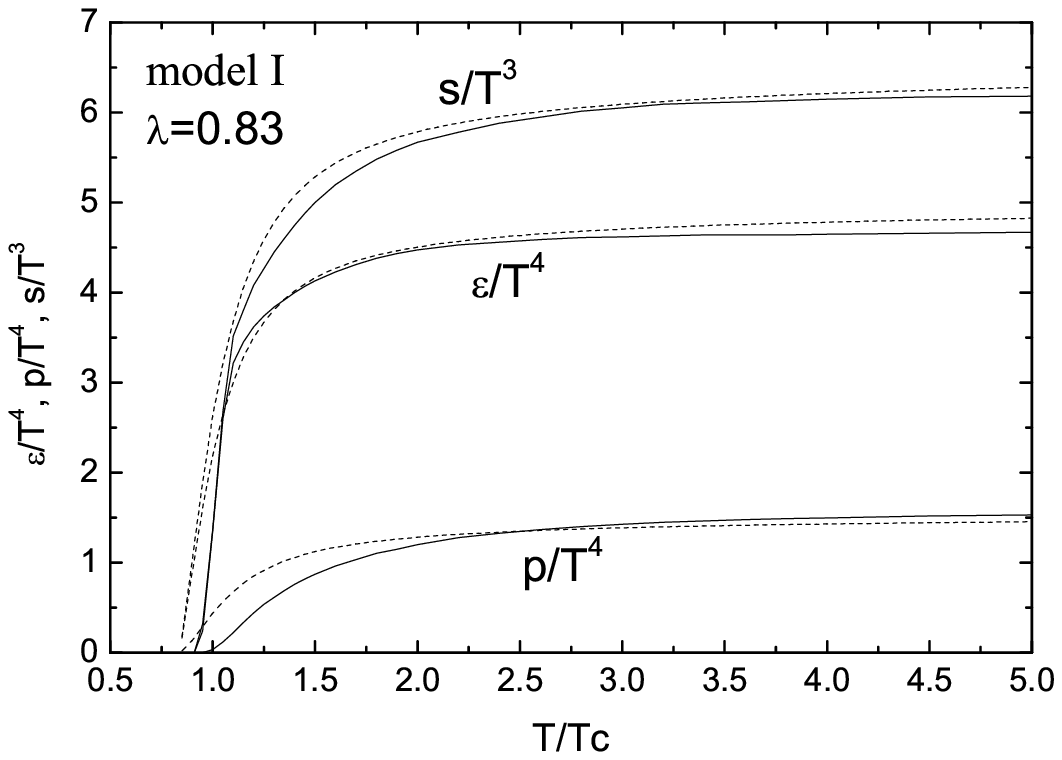}
\caption{The same as Fig.2, except that the dashed curves are the
results of treatment (III).} \label{fig5}
\end{figure}

\begin{figure}[tbp]
\includegraphics[totalheight=16cm, width=16cm]{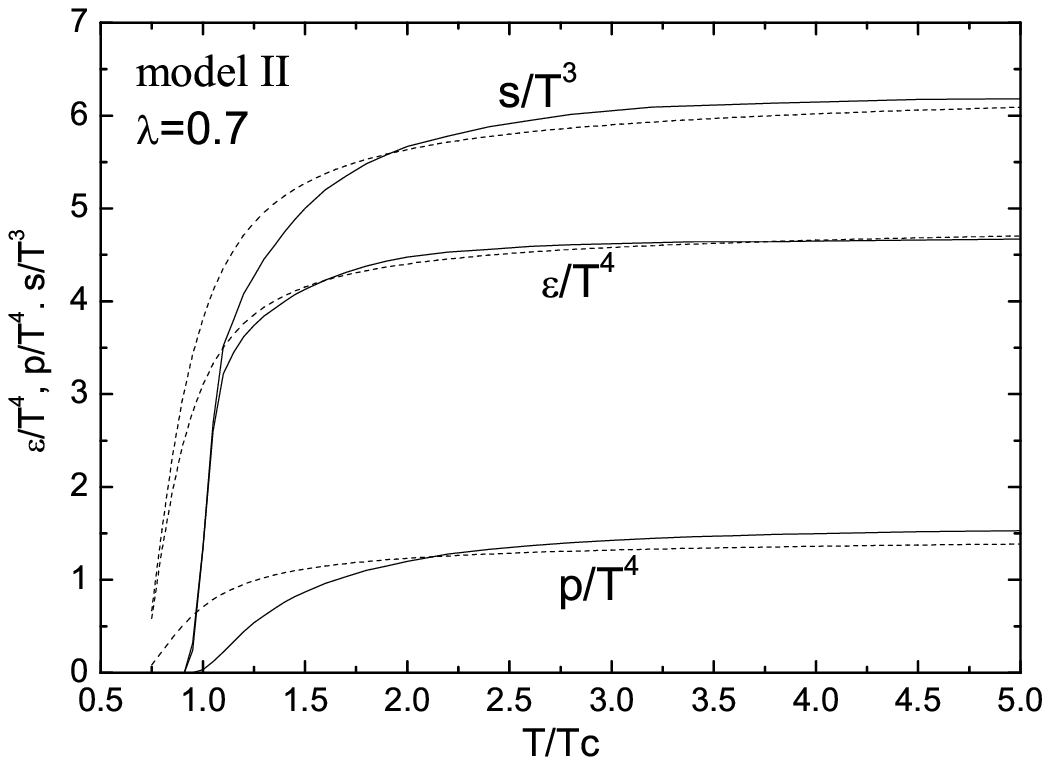}
\caption{The same as Fig.3, except that the dashed curves are the
results of treatment (III).} \label{fig6}
\end{figure}

\begin{figure}[tbp]
\includegraphics[totalheight=16cm, width=16cm]{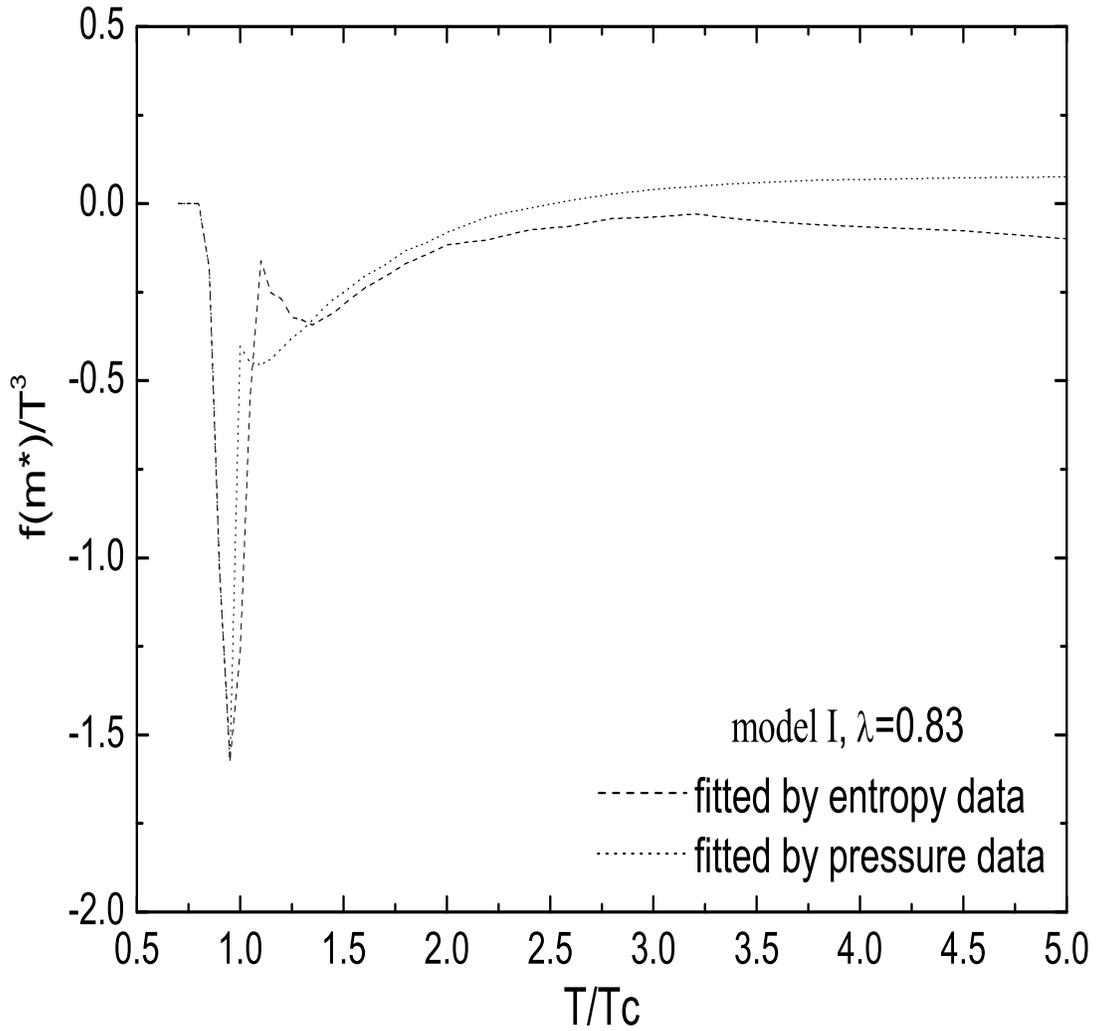}
\caption{The modification term $f(m^*)/T^3$ in SU(3) qGP model with
$m^*$ of model I and $\lambda=0.83$.} \label{fig7}
\end{figure}

\begin{figure}[tbp]
\includegraphics[totalheight=16cm, width=16cm]{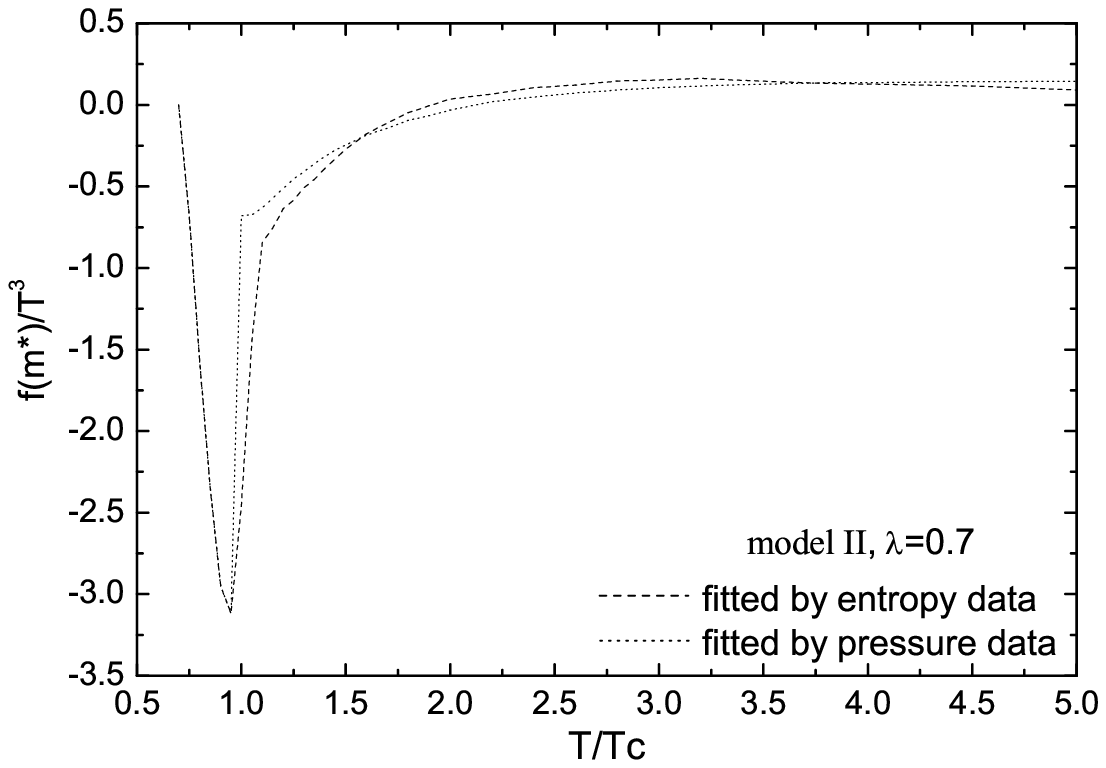}
\caption{The same as Fig.7, except that the effective mass is of
model II and $\lambda=0.7$.} \label{fig8}
\end{figure}

\begin{figure}[tbp]
\includegraphics[totalheight=16cm, width=16cm]{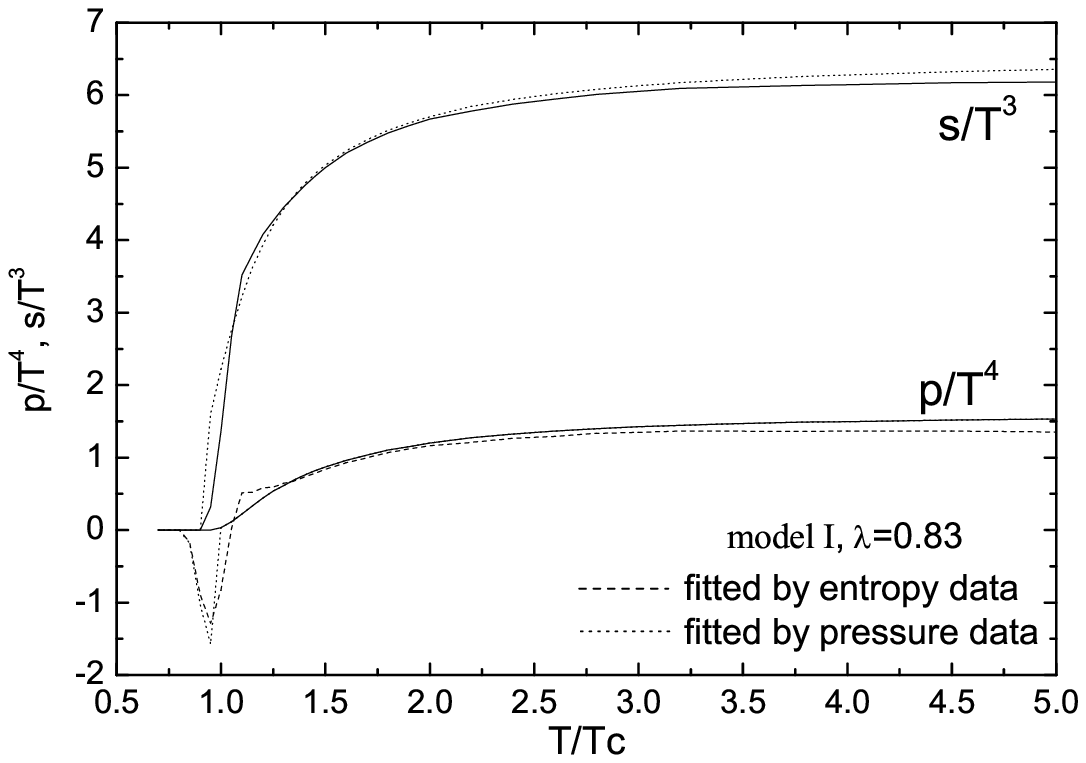}
\caption{The same as Fig.5, but the pressure and entropy density are
modified by $f(m^*)$ in Fig.7.} \label{fig9}
\end{figure}

\begin{figure}[tbp]
\includegraphics[totalheight=16cm, width=16cm]{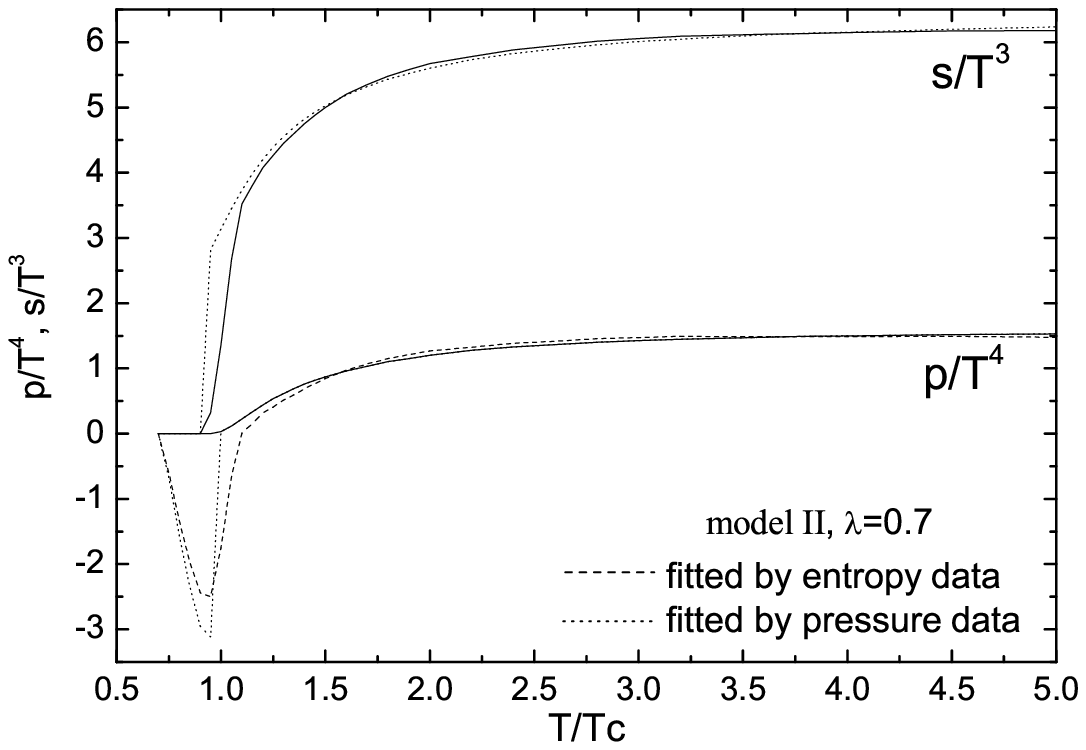}
\caption{The same as Fig.6, but the pressure and entropy density are
modified by $f(m^*)$ in Fig.8.} \label{fig10}
\end{figure}

\end{document}